\documentclass[aip,reprint,superscriptaddress]{revtex4-1}
\usepackage[USenglish]{babel} 
\usepackage[T1]{fontenc}
\usepackage{amsmath}    
\usepackage{amssymb}	
\usepackage{relsize}
\usepackage{graphicx}   
\usepackage{verbatim}   
\usepackage{color}      
\usepackage{subfigure}  
\usepackage{hyperref}   
\usepackage{soul}
\usepackage{natbib}
\usepackage{setspace}
\usepackage{placeins}
\usepackage{epstopdf}
\definecolor{light-blue}{rgb}{0.859375,0.898438,0.945317}
\definecolor{light-red}{rgb}{0.945,0.859,0.8554}
\definecolor{bright-blue}{rgb}{0,0.6875,1}
\definecolor{purple}{rgb}{.86,.4,.95}

\newcommand{\nt}[1]{{#1}}

\begin{document}

\title{Effect of lateral tip motion on multifrequency atomic force microscopy}
\author{Joseph L. Garrett}
	\address{ Department of Physics, University of Maryland, College Park, MD 20742, USA}
	\address{ Institute for Research in Electronics and Applied Physics, University of Maryland, College Park, MD 20742, USA}
\author{Lisa J. Krayer}
	\address{ Institute for Research in Electronics and Applied Physics, University of Maryland, College Park, MD 20742, USA}
	\address{ Department of Electrical and Computer Engineering, University of Maryland, College Park, MD 20742, USA}

\author{Kevin J. Palm}
	\address{ Department of Physics, University of Maryland, College Park, MD 20742, USA}
	\address{ Institute for Research in Electronics and Applied Physics, University of Maryland, College Park, MD 20742, USA}
\author{Jeremy N. Munday}	
	\address{ Institute for Research in Electronics and Applied Physics, University of Maryland, College Park, MD 20742, USA}
	\address{ Department of Electrical and Computer Engineering, University of Maryland, College Park, MD 20742, USA}

\begin{abstract}
In atomic force microscopy (AFM), the angle relative to the vertical ($\theta_{i}$) that the tip apex of a cantilever moves is determined by the tilt of the probe holder, and the geometries of the cantilever \nt{beam} and actuated eigenmode $i$. 
Even though the effects of $\theta_{i}$ on static and single-frequency AFM are known (increased effective spring constant, sensitivity to sample anisotropy, etc.), the higher eigenmodes used in multifrequency force microscopy lead to additional effects that have not been fully explored. 
Here we use Kelvin probe force microscopy (KPFM) to investigate how $\theta_{i}$ affects not only the signal amplitude and phase, but can also lead to behaviors such as destabilization of the KPFM voltage feedback loop. 
We find that longer cantilever \nt{beam}s and modified sample orientations improve voltage feedback loop stability, even though variations to scanning parameters such as shake amplitude and lift height do not.
\end{abstract}

\maketitle 
The development of specialized cantilever probes enabled atomic force microscopy (AFM)\cite{Binnig1986}. 
Later, it was realized that the holder tilts the cantilever and the trajectory of the tip apex which both increases the effective static spring constant and causes the phase of Amplitude Modulation (AM) AFM to be sensitive to both the anisotropy and slope of samples\cite{Marcus2002,DAmato2004,Heim2004,Hutter2005a}. 
For higher eigenmodes $i$, the angle between the tip apex trajectory and the vertical axis ($\theta_{i}$) also depends the geometries of the cantilever and eigenmode, so that recent experiments were able to use eigenmodes with different $\theta_{i}$ to probe forces in several directions\cite{Kawai2010a,Sigdel2013,Reiche2015,Meier2016,Huang2017,Naitoh2017}. 
Bimodal AFM, in which two eigenmodes are driven by excitation of the cantilever base, was used for most of these experiments, but it is only one of many multifrequency techniques\cite{Nonnenmacher1991,Jesse2007,Platz2008,Tetard2010,Garcia2012a,Ebeling2013b,Zerweck2005,Rajapaksa2011,Sugawara2012,Arima2015,Garrett2016,Tumkur2016,Jahng2016,Ambrosio2017}, and the effects of $\theta_{i}$ have not yet been explored for the general multifrequency case. 

\nt{Sideband multifrequency AFM methods are promising ways to investigate optoelectronic materials and devices at the nanoscale}\cite{Zerweck2005,Rajapaksa2011,Sugawara2012,Arima2015,Garrett2016,Tumkur2016,Jahng2016,Ambrosio2017}. \nt{In order to eliminate long-range artifacts and improve spatial resolution, they drive a signal by} mixing a modulated tip-sample force with piezo-driven cantilever oscillations.
\nt{A prominent sideband method is photo-induced force microscopy (PIFM), which has been used for nanoscale imaging of Raman spectra\cite{Rajapaksa2011}, nanoparticle resonances\cite{Tumkur2016}, and refractive index changes\cite{Ambrosio2017}. 
However, there is considerable debate about how to extract quantitative data from PIFM scans\cite{Jahng2016,Tumkur2016,Ambrosio2017} because it is unclear how the force couples into the probe and optical forces themselves are difficult to characterize {\it a priori}.}

\nt{Because the electrostatic force is well-characterized and controllable compared to optical forces, it offers an opportunity to test the sideband actuation technique.}
Frequency Modulation (FM) and Heterodyne (H) Kelvin probe force microscopy (KPFM) are sideband \nt{methods} that use \nt{the electrostatic force} to drive cantilever oscillations, which are in turn input into a feedback loop that measures the tip-sample potential difference. 
In a recent experiment, height variation of around 10 nm destabilized the H-KPFM voltage feedback loop, but FM-KPFM scans were stable for variations of over 100 nm\cite{Garrett2017a}. 
Because FM- and H-KPFM are primarily distinguished by the eigenmode used to amplify the KPFM signal, the cause of their qualitatively different behavior likely originates from the geometry of the eigenmodes. 
Moreover, the details of cantilever dynamics have been shown to be critical to understanding AM-KPFM\cite{Elias2011,Satzinger2012}, a much simpler technique that drives and detects its signal at a single frequency\nt{, and which can be used for comparison}.
\begin{figure}[ht]
	\centering
	\includegraphics[width=.43\textwidth]{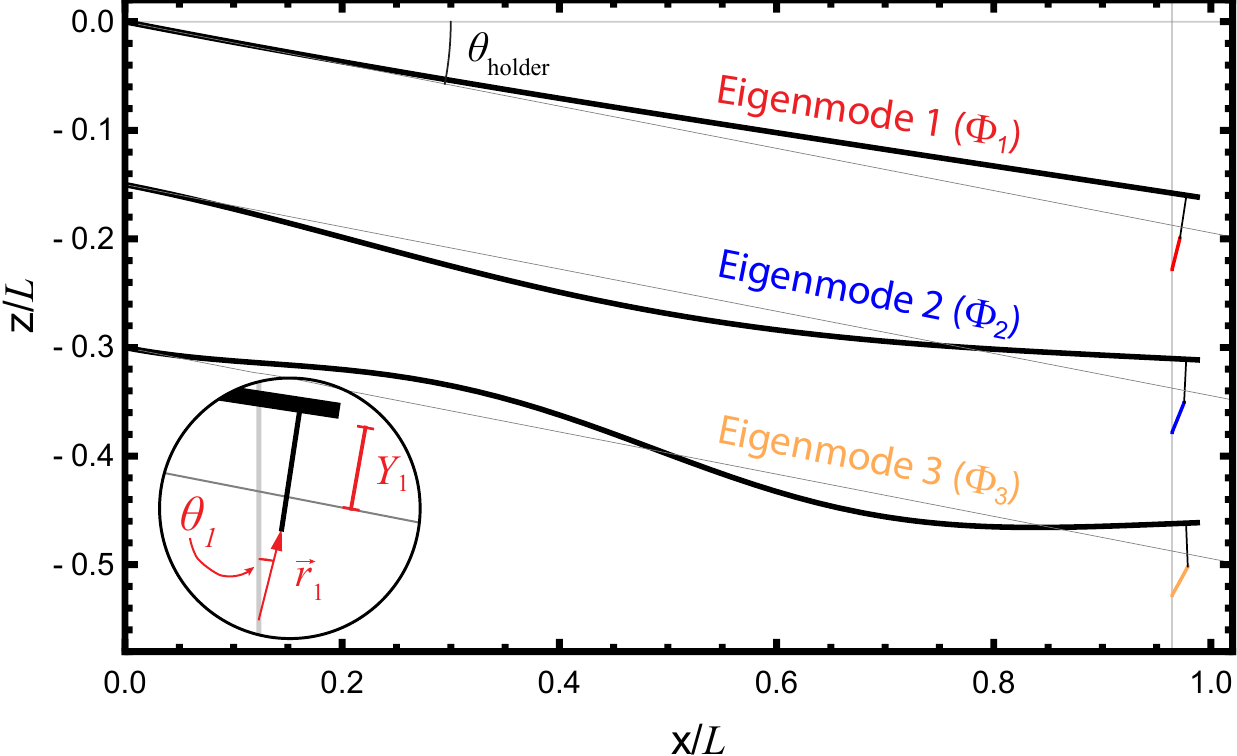}
	\caption{ The tip apex moves at an angle relative to the vertical for each eigenmode $i$ ($\theta_{i}$), which depends on the angle of the probe holder ($\theta_{\text{holder}}$), the geometry of the cantilever, and the geometry of the eigenmode ($\Phi_{i}$). 
		The inset shows the tip apex with the first eigenmode excited ($i$=1), in which the amplitude of the eigenmode ($Y_{1}$), the tip apex displacement ($\vec{r}_{1}$), and $\theta_{1}$ are labeled.
	}
	\label{fig:CantileverResponse}
\end{figure}
In this letter, we use KPFM measurements to answer the questions: (a) how does the $\theta_{i}$ of each eigenmode affect the signals of KFPM, (b) why does the KPFM feedback instability differ between H- and FM-KPFM, and (c) how \nt{do} the effects of $\theta_{i}$ \nt{appear in sideband} multifrequency force microscopy methods?

The motion of a cantilever \nt{beam} can be expressed as a sum of eigenmodes, each a solution to the Euler-Bernoulli beam equation\cite{Butt1995,Lozano2009}: 
\begin{align}
	z_{\text{cant}}(x,t) = \sum_{i=1}^{\infty} Y_{i}(t)\Phi_{i}(x),
	\label{eq:beam_motion}
\end{align}
where $Y_{i}(t)$ contains the time-dependence, $\Phi_{i}$ is the shape of the $i$th cantilever \nt{beam} eigenmode (normalized so that $\Phi_{i}(L)=1$, where $L$ is the length of the cantilever \nt{beam}), and $z_{\text{cant}}$ is the displacement of the cantilever \nt{beam} (see figure \ref{fig:CantileverResponse}). 
To maintain generality, the exact form of $\Phi_{i}$ is not specified until the numerical evaluation of $\theta_{i}$, at which point the solution for a rectangular cantilever \nt{beam} is used\cite{Butt1995,Lozano2009}. 
Thus the following analysis holds even for non-rectangular \nt{cantilever beams} and probes with large tip cones, both which may have atypical $\Phi_{i}$\cite{Tung2010,Labuda2016}.

\begin{figure}[ht]
	\centering
	\includegraphics[width=.48\textwidth]{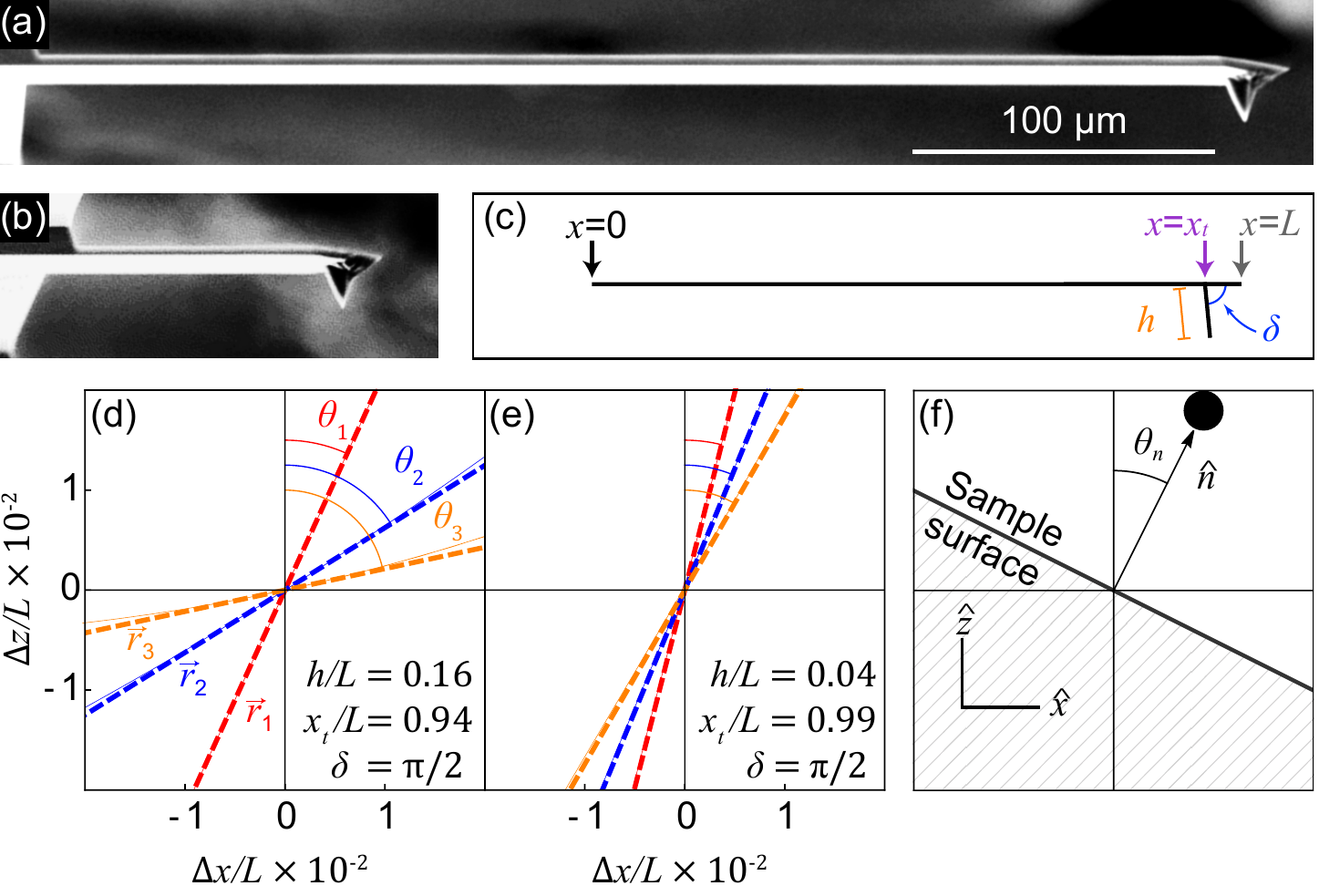}
	\caption{ Cantilever geometry determines the direction of the tip apex motion. 
		(a,b) SEM images show cantilevers of length 350 $\mu$m and 90 $\mu$m, respectively ($\mu$masch, CSC37/Pt-B and NSC35/Pt-B). 
		(c) Each cantilever is characterized by its tip cone height $h$, contact position $x_{t}$, contact angle $\delta$, and length $L$. 
		(d,e) The full calculation of $\vec{r}_{i}$ (solid line, equation \ref{eq:full_r}) and linear approximation (dashed line, equation \ref{eq:linear_r_approx}) show agreement. 
		For each eigenmode, $\theta_{i}$ is greater for the short cantilever than for the long cantilever. 
		(f) The slope of the sample is characterized by its normal vector ($\hat{n}$) and the angle it makes with the vertical ($\theta_{n}$).}
	\label{fig:Multiple_cantilevers}
\end{figure}

\nt{To calculate the trajectory of the tip apex,} the \nt{probe} is characterized by its tip cone height $h$, contact angle $\delta$, and contact position $x_{t}$ (figure \ref{fig:Multiple_cantilevers}). 
\nt{The position of the tip apex is the location of base of the tip cone \{$x_{t},Y_{i}\Phi_{i}(x_{t})$\} plus the position of the tip apex relative to the base of tip cone, \{$h\cos(\xi(Y_{i})-\delta),h\sin(\xi(Y_{i})-\delta)$\},} where $\xi(Y_{i}) = \tan^{-1}(Y_{i}\partial_{x}\Phi_{i}(x_{t}))$ is the angle of the vector normal to the cantilever at $x_{t}$.
\nt{Because the probe is held at an angle $\theta_{\text{holder}}$} (here, \nt{0.2 radians}), the displacement of the tip apex from equilibrium becomes, in the small oscillation limit ($Y_{i}\ll L$):
\begin{align}
	\vec{r}_{i} &=  \mathbf{R}\bigg[ \begin{matrix} h(\cos(\xi(Y_{i})-\delta)-\cos(\delta))  \\ Y_{i}\Phi_{i} (x_{t}) +h(\sin(\xi(Y_{i})-\delta)+\sin(\delta))  \end{matrix} \bigg],
	\label{eq:full_r}
\end{align}
where $\mathbf{R}=\left[ \begin{smallmatrix} \cos(\theta_{\text{holder}}) & \sin(\theta_{\text{holder}}) \\ -\sin(\theta_{\text{holder}}) & \cos(\theta_{\text{holder}}) \end{smallmatrix} \right]$ is a 2D rotation matrix around the base of the cantilever \nt{beam}.
For a single eigenmode in the $Y_{i}\ll L$ limit, the tip apex moves in a straight line at an angle with respect to the vertical:
\begin{align}
	 \theta_{i}= \lim_{Y_{i}/L\rightarrow 0}\cos^{-1}\big(\vec{r}_{i}\cdot(Y_{i}\hat{z})\big).
	 \label{eq:linear_r_approx}
\end{align}
\nt{Note that equations \ref{eq:full_r} and \ref{eq:linear_r_approx} imply that much of the trajectory of the tip apex is in the $\hat{x}$ direction, even for very small excitations. For example, a 10 nm amplitude excitation of the first eigenmode of the cantilever beam in figure \ref{fig:Multiple_cantilevers}b causes the tip apex to move $\approx3.9$ nm in the $\hat{x}$ direction and 8.6 nm in the $\hat{z}$ direction.}
Because the potential energy of an eigenmode must be the same whether the motion of the end of cantilever \nt{beam} ($\Phi_{i}(L)$) or the tip \nt{apex} ($\vec{r}_{i}$) is considered, an effective spring constant ($k_{i}^{\text{eff}}$) for forces acting on the tip apex parallel to $\vec{r}_{i}$ (perpendicular forces excite only eigenmodes $\neq i$) can be defined\cite{Reiche2015}:
\begin{align}
	\label{eq:sensitivity}
	k_{i}^{\text{eff}} &= \lim_{Y_{i}/L\rightarrow 0}\frac{Y_{i}^{2}}{|\vec{r}_{i}(Y_{i})|^{2}}k_{i},	
\end{align}
where $k_{i}$ is the spring constant for an upward force acting at $x=L$\cite{Melcher2007}.

\begin{figure}[hb]
	\centering
	\includegraphics[width=.43\textwidth]{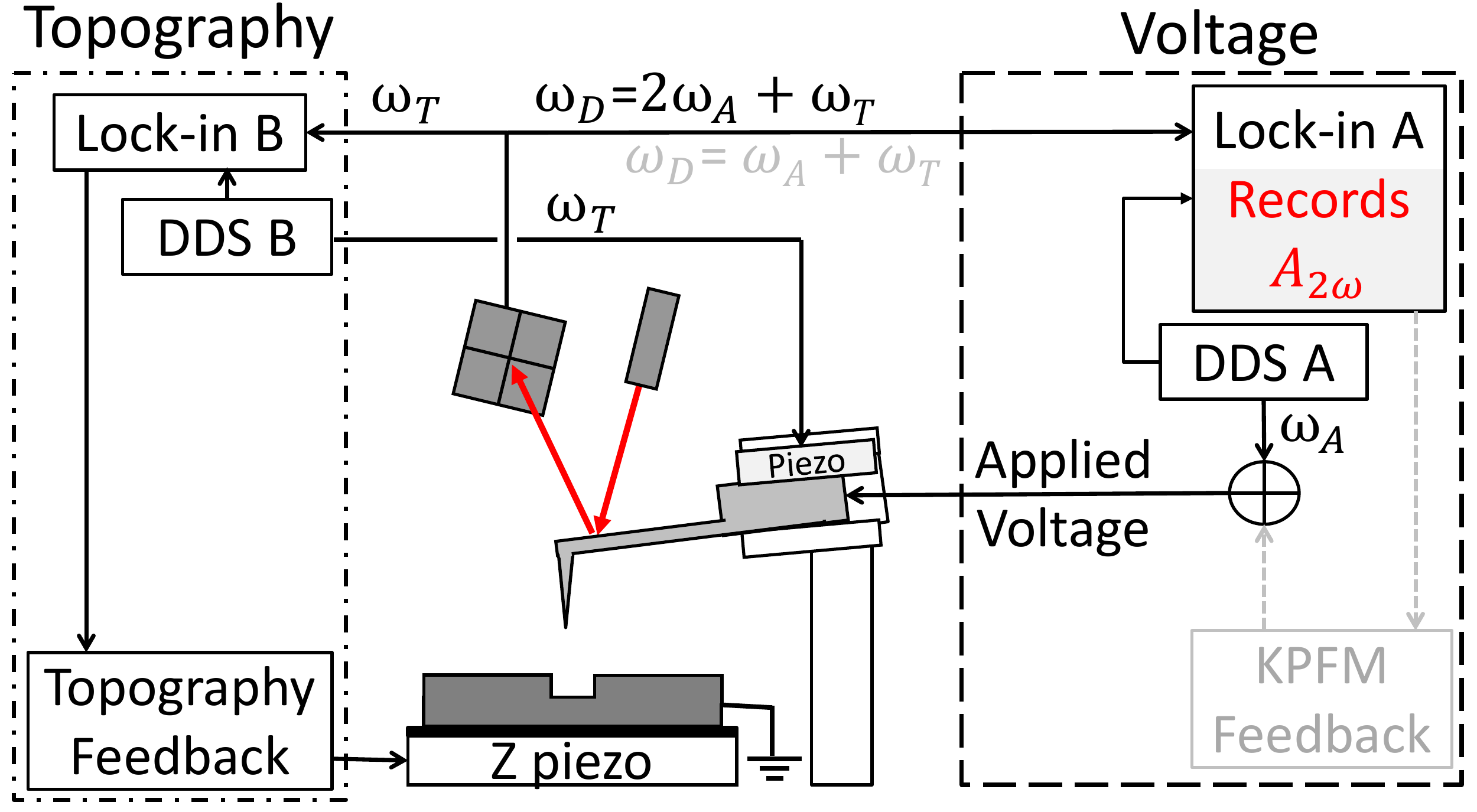}
	\caption{ 
	\nt{An AC voltage, $V_{AC}$, is applied to the \nt{cantilever} at frequency $\omega_{A}$, while tip-sample separation is controlled by piezo-driven oscillation at frequency $\omega_{T}$ and the sample is grounded. The oscillations at $\omega_{T}$ mix with the electrostatic force driven by $V_{AC}$ at frequency $\omega_{A}$ to drive the tip apex at the detection frequency, $\omega_{D}$, which is amplified by one of the cantilever's resonance frequencies and detected by a lock-in amplifier. When KPFM feedback is used, the grey signal paths are added to the circuit, and $\omega_{A}=(\omega_{D}-\omega_{T})/2$ is changed to $\omega_{A}=\omega_{D}-\omega_{T}$.}
	}
	\label{fig:BoxDiagram}
\end{figure}

\begin{figure*}[ht]
	\centering
	\includegraphics[width=.90\textwidth]{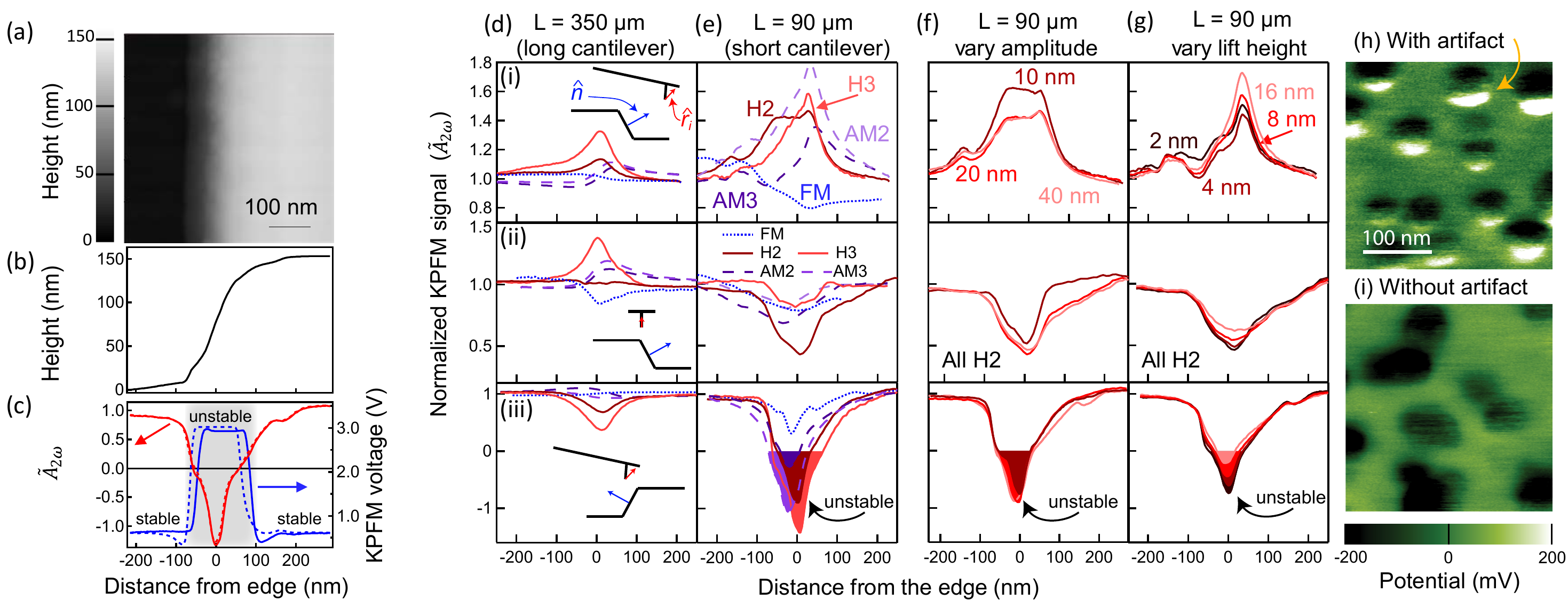}
	\caption{
		(a,b) The height of a trench that is scanned with KPFM ($128^{2}$ pixels, 500 nm/s). 
		(c) Where the normalized \nt{signal} ($\tilde{A}_{2\omega}$, red) becomes negative, the KPFM voltage with feedback on (blue) becomes unstable and approaches the limit imposed on the feedback loop, for both trace (solid) and retrace (dashed). 
		Long (d) and short (e-g) cantilever \nt{beam}s scan across the trench edge in three different orientations: down (i, $\theta_{n}<0$), parallel to (ii), and up the slope (iii, $\theta_{n}>0$). 
		In (i,ii), $\tilde{A}_{2\omega}$ remains positive for all methods, but in (iii) all methods except FM-KPFM contain a negative portion for the short cantilever \nt{beam}.  
		(f,g) Varying scan parameters such as $A_{T}$ (used for topography control) and lift height are not sufficient to prevent $\tilde{A}_{2\omega}$ < 0.
		\nt{The different scanning modes are labeled by a prefix (eg. `H' for H-KPFM) and a number indicating the eigenmode used to amplify the signal, except for FM-KPFM, which always uses the first eigenmode. (h) An artifact is present in an H-KFPM scan of Au nanoparticles on indium tin oxide when the signal is amplified by the second eigenmode of the short cantilever \nt{beam}. (i) When the long cantilever \nt{beam} is used to scan other Au nanopaticles on the same sample, the artifact is eliminated.}
	}
	\label{fig:Signals_and_geometry}
\end{figure*}

\begin{table}[b]
	\caption{Cantilever resonance frequencies}
	\centering
	\begin{tabular}{ c | c  c  c  c  c  c  c   }
		
		L ($\mu$m) & $\frac{\omega_{1}}{2\pi}$  (MHz)& $\frac{\omega_{2}}{2\pi}$ & \hspace{5 pt}$\frac{\omega_{3}}{2\pi}$\hspace{5 pt} &  $\frac{\omega_{4}}{2\pi}$ & \hspace{5 pt}$\frac{\omega_{5}}{2\pi}$\hspace{5 pt}  & \hspace{5 pt}$\frac{\omega_{6}}{2\pi}$\hspace{5 pt} & \hspace{5 pt}$\frac{\omega_{7}}{2\pi}$\hspace{5 pt}  \\
		
		\hline
		
		90 & 0.25 & 1.62 & 4.58 & - & - & - & - \\
		350 & 0.02& 0.13& 0.37& 0.72& 1.20& 1.79& 2.50\\
	
	\end{tabular}
	
	\label{table:cantilever_resonances}
\end{table}

The tip \nt{apex trajectory} affects \nt{{AFM}} techniques \nt{that} use a modulated tip-sample force $\vec{F}_{\text{dir}}$ to actuate the cantilever either directly or through sideband coupling while relying on \nt{piezo-driven oscillation with amplitude $A_{T}$} at frequency $\omega_{T}$ for topography control \nt{(here, $\omega_{T}=\omega_{1}$ in table \ref{table:cantilever_resonances} is used)}.
\nt{Sideband techniques generate a signal by modulating a separation-dependent force $\vec{F}_{\text{dir}}$ at frequency $\omega_{M}$, which is then mixed with the piezo-driven oscillations, typically $A_{T}$. Here, the resonance frequency used for detection determines the modulation frequency $\omega_{M} = \omega_{i}-\omega_{T}$ (table \ref{table:cantilever_resonances}).
By using the force gradient, sideband methods exclude the non-local effects of the cantilever \nt{beam} that are present when $\vec{F}_{\text{dir}}$ is used for direct actuation, such as in AM-KPFM\cite{Zerweck2005,Sugawara2012,Jahng2016}.

To confirm that the cantilever beam's contribution to the total force is small even when higher eigenmodes are used, the force on  the beam is computed for both direct actuation ($-\partial U/ \partial Y_{i}$) and sideband actuation ($-\partial^{2} U/ \partial Y_{i}^{2}$), where $U$ is the electrostatic potential energy between the probe and the surface evaluated using the proximity force approximation and the geometry of the longer probe. The contribution from the tip apex is calculated by modeling it as a 30 nm radius sphere 10 nm above the surface. The percent of the signal originating from the cantilever beam using direct actuation is found to be 17-53\% for the first seven eigenmodes, while with sideband actuation 0.1-0.2\% of the signal originates from the beam. The small contribution from the beam validates the approximation that the electrostatic force acts on the tip apex for sideband actuation of higher eigenmodes.} 

In the small-oscillation approximation\cite{Jahng2016,Garrett2016}, the force driving sideband oscillation \nt{is $\vec{F}_{\text{side}}\cos(\omega_{D}t)$, where}
\begin{align}
\label{eq:SidebandActuation}
\vec{F}_{\text{side}} = \partial_{d}\vec{F}_{\text{dir}}\frac{A_{T}}{2}\cos(\theta_{i}-\theta_{n}),
\end{align}
\nt{in which} $d$ is the tip-sample separation, \nt{$\omega_{D}$ is the detection frequency}, and the $\cos(\theta_{i}-\theta_{n})$ factor originates from the angle between the trajectory \nt{of the tip apex} and the force vector (parallel to $\hat{n}$).
The \nt{displacement of the tip apex at $\omega_{D}$ is then $\vec{r}_{j}(t)=A_{D}\cos(\omega_{D}t)\hat{r}_{j}$, where eigenmode $j$ is driven and the signal detected by the lock-in amplifier is}
\begin{align}
	A_{D} &= \frac{Q_{j}}{k^{\text{eff}}_{j}}\vec{F}\cdot\hat{r}_{j},	
	\label{eq:direct}
\end{align}
for both the sideband and direct driving forces \nt{(figure \ref{fig:BoxDiagram}). A change in the sign of $A_{D}$ corresponds to a phase shift by $\pi$ radians.}

The interplay of $\theta_{j}$ and sample slope can then be observed in the \nt{signal $A_{D}$} normalized by the \nt{its value} on a flat surface ($\tilde{A}_{D} \equiv \frac{A_{D}}{A_{D}(\theta_{n}=0)}$):

\begin{align}
	\label{eq:single_freq_A}
	\tilde{A}_{D}^{\text{dir}} &= \frac{\cos(\theta_{j}-\theta_{n})}{\cos(\theta_{j})},\\
	\tilde{A}_{D}^{\text{side}} &=\frac{\cos(\theta_{n}-\theta_{j})\cos(\theta_{n}-\theta_{i})}{\cos(\theta_{j})\cos(\theta_{i})},
	\label{eq:mult_freq_A}
\end{align}
where it is assumed that $\hat{n}$ is in the x-z plane and $\theta_{i},\theta_{j}\neq\pm\pi/2$.
Note that if $|\theta_{i}-\theta_{n}|>\frac{\pi}{2}>\theta_{i}$, $\tilde{A}_{D}$ changes sign.

Equations \ref{eq:single_freq_A} and \ref{eq:mult_freq_A} predict how the geometry of tip apex motion causes scanning probe methods to be sensitive to sample slope. 
To test the equations, a silicon trench is fabricated using e-beam lithography to pattern a 2 $\mu$m $\times$ 100 $\mu$m line on a silicon wafer which is then etched using reactive ion etching (RIE) and coated with 5 nm of chromium for conductivity. 
The edges of the trench are imaged, in attractive mode \cite{Paulo2002}, (Cypher, Asylum Research), trace and retrace images are averaged, and each column of pixels is summed and averaged (figure \ref{fig:Signals_and_geometry}a,b). 

\nt{In the static limit,} when an AC voltage is applied to a probe \nt{at frequency $\omega_{A}$}, the tip-sample electrostatic force has components at three frequencies\cite{Nonnenmacher1991,Zerweck2005}: $\vec{F}_{\text{es}} = \vec{F}_{DC} + \vec{F}_{\omega}\cos(\omega_{A} t) + \vec{F}_{2\omega}\cos(2\omega_{A} t).$
\nt{Either $\vec{F}_{\omega}$ or $\vec{F}_{2\omega}$ can be used in equation \ref{eq:SidebandActuation} to drive the sideband signal by choosing $\omega_{M} = \omega_{A}$ or $2\omega_{A}$, respectively.
The signal then depends on the gradient of the original modulation force\cite{Zerweck2005,Sugawara2012,Miyahara2017}. For FM-KPFM, $\omega_{A}\ll\omega_{1}$\cite{Zerweck2005}.}
Closed loop KPFM measures the contact potential difference between the probe and sample using a feedback loop to nullify a signal driven by the force $\vec{F}_{\omega}$.
Alternatively, open loop KPFM uses oscillation driven by $\vec{F}_{2\omega}$ combined with the $\vec{F}_{\omega}$ signal to estimate the potential difference \nt{$\Delta V$ from the relationship between the forces $\vec{F}_{2\omega} = \vec{F}_{\omega}V_{AC}/(4\Delta V)$}\cite{Takeuchi2007, Collins2015}. 
The relationship between $\vec{F}_{2\omega}$ \nt{(which drives $A_{2\omega}$ according to equation \ref{eq:direct})} and KPFM feedback loop \nt{itself} can be seen in figure \ref{fig:Signals_and_geometry}c: the feedback becomes unstable at locations where $A_{2\omega}$ changes sign.
Moreover, any change in $A_{2\omega}$ makes KPFM susceptible to topographic cross-talk\cite{Barbet2014}.  
The \nt{signal} is driven by $\vec{F}_{2\omega}$ because it reveals the behavior of the KPFM feedback loop, without requiring feedback to be used and is not susceptible to patch potentials or tip change. 

The effect of slope is revealed by observing how the normalized signal ($\tilde{A}_{2\omega}$) changes as the \nt{tip apex} approaches an edge of the trench at different orientations, for AM-, FM- and H-KPFM with the first three eigenmodes of each cantilever\nt{, and $V_{AC}$ = 3 V}.
In figure \ref{fig:Signals_and_geometry} the trench edge is crossed with three different orientations: (i) the vector from the base of the cantilever \nt{beam} to its tip \nt{apex} points down the slope ($\theta_{n}>0$, from the higher to the lower level) (ii) parallel to the slope ($\hat{n}$ out of plane) and (iii) up the slope ($\theta_{n}<0$). 
One trend predicted by equation \ref{eq:mult_freq_A} is observed: $\tilde{A}_{2\omega}$ tends to increase as $\theta_{n}$ increases.  
However, the decrease of $\tilde{A}_{2\omega}$ is greater for the short cantilever \nt{beam} than for the long cantilever \nt{beam}. 
For the short cantilever \nt{beam}, the $\theta_{n}<0$ edge leads to $\tilde{A}_{2\omega} < 0$ for every technique except FM-KPFM.
	
Other scan parameters affect $\tilde{A}_{2\omega}$ much less.
$A_{T}$, used for topography control, is varied from 10 to 40 nm, but the shape of $\tilde{A}_{2\omega}$ retains a negative portion as the $\theta_{n}<0$ edge is crossed. 
Similarly, using a two-pass method and varying the lift height from 2 nm to 16 nm does not prevent $\tilde{A}_{2\omega}<0$ at the $\theta_{n}<0$ edge.
Thus, if KPFM feedback is unstable for geometric reasons, adjustments to the scan settings do not typically stablize it.
		
\begin{figure}[ht]
	\centering
	\includegraphics[width=.44\textwidth]{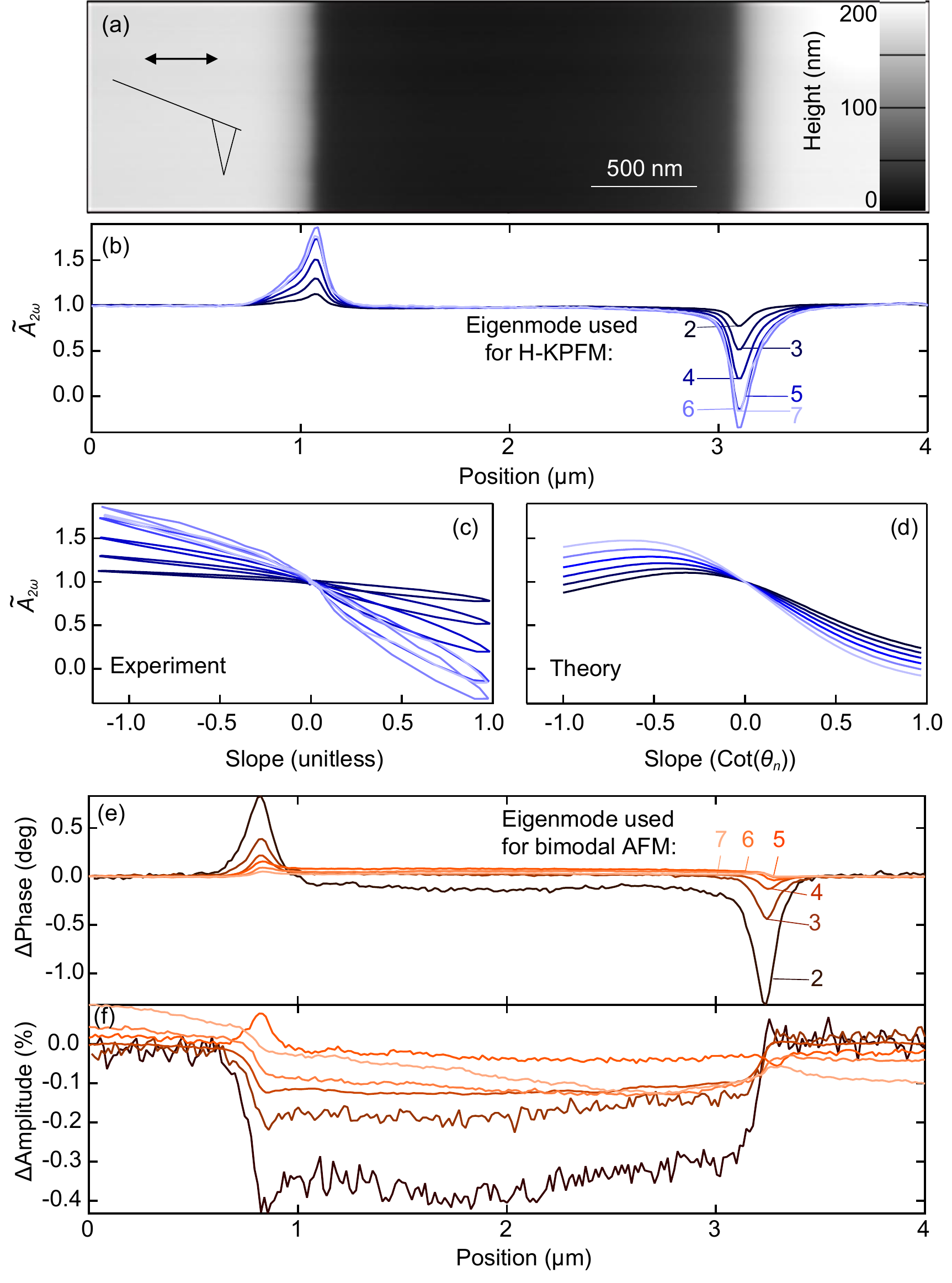}
	\caption{ (a) A cantilever scans, using H-KPFM, across a 2 $\mu$m wide chromium-coated silicon trench (64$\times$256 pixels, 800 nm/s).
			(b) On the downward slope (left), the normalized signal $\tilde{A}_{2\omega}$ becomes larger, but 
			on the upward slope (right), the signal decreases. 
			(c) Measured and (d) predicted values of $\tilde{A}_{2\omega}$ are plotted against the local slope of the trench.
			Higher eigenmodes tend to show a greater change with slope, as predicted from their larger $\theta_{i}$. 
			Bimodal AFM is also used to scan across the surface, \nt{while biased to 3 V}. 
			(e) The change in phase shows peaks at the edges, but unlike the H-KPFM case, the relative amplitude of the phase change decreases for higher eigenmodes, because the of increased $k^{\text{eff}}_{i}$.
			(f) The amplitude decreases in the middle of the trench, but not at the edges, and changes by $<0.5\%$.
			}
	\label{fig:Multiple_eigenmodes}
\end{figure}
	
To test the predictions with a wider range of $\theta_{i}$, the trenches are scanned again with the long probe in H-KPFM mode using the first eigenmode for topography control and amplifying the $\vec{F}_{2\omega}$ signal with eigenmodes 2-7 \nt{(ie. $\omega_{A}=\omega_{M}/2 = (\omega_{i}-\omega_{1})/2$, so that $\omega_{D}=\omega_{i}$ for 2$\leq i\leq$7, table \ref{table:cantilever_resonances})}. 
Because each eigenmode has a slightly greater $\theta_{i}$ than the one before it (ie. $\theta_{i+1}>\theta_{i}$), equation \ref{eq:mult_freq_A} predicts that the effect of sample slope is greater for the higher eigenmodes than the lower ones, and the experiment confirms this trend, although the seventh eigenmode changes less than the sixth (figure \ref{fig:Multiple_eigenmodes}b-d). 
The experimental data do not all fall on a single line (figure \ref{fig:Multiple_eigenmodes}c), perhaps because the region on the sample from which the $\vec{F}_{2\omega}$ force originates deviates from the single-slope assumption.
\nt{For eigenmodes 3-7, the data agree better with equation \ref{eq:mult_freq_A}, which has no free parameters, than with the null hypothesis that the signal does not depend on slope, thus confirming that the direction of the force affects how it drives the tip apex. However, equation \ref{eq:mult_freq_A} tends to underestimate $\tilde{A}_{2\omega}$, particularly for slopes $<-0.5$, which suggests that other factors, such as the tip cone and changes to the piezo-driven oscillation, $A_{T}$, may also matter. 
An initial test of effect of slope on piezo-driven oscillation with bimodal AFM shows a change in the phase at the edges of the trench (figure \ref{fig:Multiple_eigenmodes}e,f). Because the sideband excitation technique is similar for different forces, the results here indicate that $\theta_{i}$ affects the whole class of methods.}

The direction of the tip \nt{apex} trajectory depends on cantilever geometry and the eigenmodes used, and influences \nt{sideband} multifrequency force microscopy methods. 
It can even change the sign of the signal, which leads to feedback instability in KPFM. 
The results here show that considerable topographic restrictions exist for multifrequency methods when short cantilevers are used.
Because short cantilevers enable faster scanning than long cantilevers\cite{Walters1996a}, the restriction amounts to a speed limitation for any given roughness.
Because the equations above separate the calculation of $\theta_{i}$ (\ref{eq:beam_motion}-\ref{eq:sensitivity}) from the analysis of \nt{the sideband signal} (\ref{eq:SidebandActuation}-\ref{eq:mult_freq_A}),
either portion can be combined with numerical methods to account for non-rectangular cantilever \nt{beams}, or non-analytic forces.
Knowledge of the effect of geometry will assist in the development of \nt{additional} multifrequency methods and will make the interpretation of current methods more accurate. In particular, the improved stability of KPFM will enable high resolution voltage mapping of rough or textured surfaces, which will allow for improved nanoscale characterization of optoelectronic structures such as solar cells and for the study of light induced charging effects resulting from hot carrier generation or plasmoelectric excitation of nanostructured metals\cite{Sheldon2014,Tumkur2016,Garrett2017a}.

The authors acknowledge funding support from the Office of Naval Research Young Investigator Program (YIP) under Grant No. N00014-16-1-2540, and the support of the Maryland NanoCenter and its FabLab. LK acknowledges that this material is based upon work supported by the National Science Foundation Graduate Research Fellowship under DGE 1322106. 

\bibliographystyle{iopart_num_edited}
\bibliography{AFM_geometry}

\providecommand{\newblock}{}
\begin{thebibliography}{10}
\expandafter\ifx\csname url\endcsname\relax
  \def\url#1{{\tt #1}}\fi
\expandafter\ifx\csname urlprefix\endcsname\relax\def\urlprefix{URL }\fi
\providecommand{\eprint}[2][]{\url{#2}}

\bibitem{Binnig1986}
Binnig G, Quate C and Gerber C 1986 {\em Phys. Rev. Lett.\/} {\bf 56}

\bibitem{Marcus2002}
Marcus M~S, Carpick R~W, Sasaki D~Y and Eriksson M~A 2002 {\em Phys. Rev.
  Lett.\/} {\bf 88} 226103

\bibitem{DAmato2004}
D'Amato M~J, Marcus M~S, Eriksson M~A and Carpick R~W 2004 {\em Appl. Phys.
  Lett.\/} {\bf 85} 4738--4740

\bibitem{Heim2004}
Heim L~O, Kappl M and Butt H~J 2004 {\em Langmuir\/} {\bf 20} 2760--2764

\bibitem{Hutter2005a}
Hutter J~L 2005 {\em Langmuir\/} {\bf 21} 2630--2

\bibitem{Kawai2010a}
Kawai S, Glatzel T, Koch S, Such B, Baratoff A and Meyer E 2010 {\em Phys. Rev.
  B\/} {\bf 81} 085420

\bibitem{Sigdel2013}
Sigdel K~P, Grayer J~S and King G~M 2013 {\em Nano Lett.\/} {\bf 13} 5106--5111

\bibitem{Reiche2015}
Reiche C~F, Vock S, Neu V, Schultz L, B{\"{u}}chner B and M{\"{u}}hl T 2015
  {\em New J. Phys.\/} {\bf 17} 013014

\bibitem{Meier2016}
Meier T, Eslami B and Solares S~D 2016 {\em Nanotechnology\/} {\bf 27} 085702

\bibitem{Huang2017}
Huang F, Tamma V~A, Rajaei M, Almajhadi M and Wickramasinghe H~K 2017 {\em
  Appl. Phys. Lett.\/} {\bf 110} 063103

\bibitem{Naitoh2017}
Naitoh Y, Turansk{\'{y}} R, Brndiar J, Li Y~J, {\v{S}}tich I and Sugawara Y
  2017 {\em Nat. Phys.\/} {\bf 13} 663--667

\bibitem{Nonnenmacher1991}
Nonnenmacher M, O'Boyle M~P and Wickramasinghe H~K 1991 {\em Appl. Phys.
  Lett.\/} {\bf 58} 2921

\bibitem{Jesse2007}
Jesse S, Kalinin S~V, Proksch R, Baddorf A~P and Rodriguez B~J 2007 {\em
  Nanotechnology\/} {\bf 18} 435503

\bibitem{Platz2008}
Platz D, Thol{\'{e}}n E~A, Pesen D and Haviland D~B 2008 {\em Appl. Phys.
  Lett.\/} {\bf 92} 153106

\bibitem{Tetard2010}
Tetard L, Passian A and Thundat T 2010 {\em Nat. Nanotechnol.\/} {\bf 5}
  105--109

\bibitem{Garcia2012a}
Garcia R and Herruzo E~T 2012 {\em Nat. Nanotechnol.\/} {\bf 7} 217--26

\bibitem{Ebeling2013b}
Ebeling D, Eslami B and Solares S~D~J 2013 {\em ACS Nano\/} {\bf 7}
  10387--10396

\bibitem{Zerweck2005}
Zerweck U, Loppacher C, Otto T, Grafstr{\"{o}}m S and Eng L~M 2005 {\em Phys.
  Rev. B\/} {\bf 71} 125424

\bibitem{Rajapaksa2011}
Rajapaksa I and {Kumar Wickramasinghe} H 2011 {\em Appl. Phys. Lett.\/} {\bf
  99} 161103

\bibitem{Sugawara2012}
Sugawara Y, Kou L, Ma Z, Kamijo T, Naitoh Y and {Jun Li} Y 2012 {\em Appl.
  Phys. Lett.\/} {\bf 100} 223104

\bibitem{Arima2015}
Arima E, Naitoh Y, {Jun Li} Y, Yoshimura S, Saito H, Nomura H, Nakatani R and
  Sugawara Y 2015 {\em Nanotechnology\/} {\bf 26} 125701

\bibitem{Garrett2016}
Garrett J~L and Munday J~N 2016 {\em Nanotechnology\/} {\bf 27} 245705

\bibitem{Tumkur2016}
Tumkur T~U, Yang X, Cerjan B, Halas N~J, Nordlander P and Thomann I 2016 {\em
  Nano Lett.\/} {\bf 16} 7942--7949

\bibitem{Jahng2016}
Jahng J, Kim B, Lee E~S and Potma E~O 2016 {\em Phys. Rev. B\/} {\bf 94} 195407

\bibitem{Ambrosio2017}
Ambrosio A, Devlin R~C, Capasso F and Wilson W~L 2017 {\em ACS Photonics\/}
  {\bf 4} 846--851

\bibitem{Garrett2017a}
Garrett J~L, Tennyson E~M, Hu M, Huang J, Munday J~N and Leite M~S 2017 {\em
  Nano Lett.\/} {\bf 17} 2554--2560

\bibitem{Elias2011}
Elias G, Glatzel T, Meyer E, Schwarzman A, Boag A and Rosenwaks Y 2011 {\em
  Beilstein J. Nanotechnol.\/} {\bf 2} 252--260

\bibitem{Satzinger2012}
Satzinger K~J, Brown K~A and Westervelt R~M 2012 {\em J. Appl. Phys.\/} {\bf
  112} 064510

\bibitem{Butt1995}
Butt H~J and Jaschke M 1995 {\em Nanotechnology\/} {\bf 6} 1--7

\bibitem{Lozano2009}
Lozano J and Garcia R 2009 {\em Phys. Rev. B\/} {\bf 79} 014110

\bibitem{Tung2010}
Tung R~C, Wutscher T, Martinez-Martin D, Reifenberger R~G, Giessibl F and Raman
  A 2010 {\em J. Appl. Phys.\/} {\bf 107} 104508

\bibitem{Labuda2016}
Labuda A, Kocun M, Lysy M, Walsh T, Meinhold J, Proksch T, Meinhold W, Anderson
  C and Proksch R 2016 {\em Rev. Sci. Instrum.\/} {\bf 87} 073705

\bibitem{Melcher2007}
Melcher J, Hu S and Raman A 2007 {\em Appl. Phys. Lett.\/} {\bf 91} 053101

\bibitem{Paulo2002}
Paulo {\'{A}} and Garc{\'{i}}a R 2002 {\em Phys. Rev. B\/} {\bf 66} 041406

\bibitem{Miyahara2017}
Miyahara Y and Grutter P 2017 {\em Appl. Phys. Lett.\/} {\bf 110} 163103

\bibitem{Takeuchi2007}
Takeuchi O, Ohrai Y, Yoshida S and Shigekawa H 2007 {\em Jpn. J. Appl. Phys.\/}
  {\bf 46} 5626--5630

\bibitem{Collins2015}
Collins L, Jesse S, Balke N, Rodriguez B~J, Kalinin S and Li Q 2015 {\em Appl.
  Phys. Lett.\/} {\bf 106} 104102

\bibitem{Barbet2014}
Barbet S, Popoff M, Diesinger H, Deresmes D, Th{\'{e}}ron D and M{\'{e}}lin T
  2014 {\em J. Appl. Phys.\/} {\bf 115} 144313

\bibitem{Walters1996a}
Walters D~A, Cleveland J~P, Thomson N~H, Hansma P~K, Wendman M~A, Gurley G and
  Elings V 1996 {\em Rev. Sci. Instrum.\/} {\bf 67} 3583--3590

\bibitem{Sheldon2014}
Sheldon M~T, Groep J~V~D, Brown A~M, Polman A and Atwater H~A 2014 {\em
  Science\/} {\bf 346} 828--831

\end{thebibliography}

\end{document}